\documentclass[12pt]{article}

\usepackage[
	includefoot,
	right=1in,
	left=1in, 
	top=1in, 
	bottom=.8in, 
	headsep=.5in, 
	footskip=.7in,
]{geometry}
\usepackage{bm}
\usepackage{float}
\usepackage{dcolumn}
\usepackage{graphicx}
\usepackage{caption}
\usepackage{subcaption}
\usepackage{chngcntr}
\usepackage{hyperref}
\usepackage[affil-it]{authblk}
\usepackage{abstract}
\usepackage[pagestyles]{titlesec}
\usepackage{titletoc}
\usepackage{amssymb}
\usepackage{amsmath}
\usepackage{bbold}
\usepackage[sorting=none]{biblatex}
\usepackage{sidecap}
\usepackage{cancel}
\usepackage{comment}
\usepackage[pagewise]{lineno}
\usepackage[dvipsnames]{xcolor}
\usepackage[math]{cellspace}
\usepackage{setspace}
\usepackage{ragged2e}

\titlelabel{\thetitle\hspace{1ex}}

\renewcommand{\thesubsection}{\arabic{section}.\arabic{subsection}}

\titleformat*{\section}{\sf\large\bfseries}
\titlespacing*{\section}{0pt}{1.6em}{.8em}

\titleformat{\subsection}{\sf\normalsize\bfseries}{\thesubsection}{1ex}{}
\titlespacing*{\subsubsection}{0pt}{.8em}{.8em}

\newpagestyle{main}{ \sethead{}{}{} \setfoot{}{}{\ifthesection{\thesection}{}\hspace{1ex}{\itshape\sectiontitle}\qquad\thepage} }
\renewpagestyle{plain}{ \setfoot{}{}{\thepage} }
\titlecontents{section}[0em]
    {}
    {\thecontentslabel\hspace{1ex}}
    {}
    {\hspace{1ex}\bfseries\thecontentspage}
\titlecontents{subsection}[1.8em]
    {}
    {\thecontentslabel\hspace{1ex}}
    {}
    {\hspace{1ex}\bfseries\thecontentspage}
    
\newcommand{\x}[1]{\text{#1}} 
\newcommand{\id}{\, \x{d}} 
\renewcommand{\d}{\x{d}} 
\newcommand{\Lagr}{\mathcal{L}}
\renewcommand{\L}{\Lagr}

\newcommand{\GG}{\left(\frac{1}{4}F_{\alpha \beta} \tilde{F}^{\alpha \beta}\right)}
\newcommand{\GGmu}{\left(\frac{1}{4}F_{\mu \nu} \tilde{F}^{\mu \nu}\right)}

\def\dell{\partial}
\newcommand{\be}{\begin{equation}}
\newcommand{\ee}{\end{equation}}
\def\bea{\begin{eqnarray}}
\def\eea{\end{eqnarray}}
\newcommand{\eq}[1]{(\ref{#1})}
\newcommand{\rhat}{\mathbf{\hat{r}}}

\newcommand{\ordr}[2]{\mathcal{O}(#1^{#2})}
\newcommand{\ordrb}[1]{\ordr{b}{-#1}}
\newcommand{\ordrep}[1]{\ordr{\epsilon}{#1}}
\newcommand{\usol}[1]{u^{(#1)}}
\newcommand{\conj}[1]{\quad \x{#1} \quad}
\def\where{\conj{where}}
\def\with{\conj{with}}
\def\and{\conj{and}}
\newcommand{\ol}[1]{\mkern 1.5mu\overline{\mkern-1.5mu{#1}\mkern-1.5mu}\mkern 1.5mu}
\newcommand{\partialfrac}[2]{\frac{\partial #1}{\partial #2}}
\newcommand{\dpartialfrac}[2]{\dfrac{\partial #1}{\partial #2}}
\newcommand{\fpartialfrac}[2]{\partial #1 / \partial #2}
\def\zbar{\ol{z}}
\def\bi{\textsc{bi}}

\DeclareMathOperator{\tr}{tr}

\def\nn{\nonumber}

\begin{document}

\thispagestyle{plain}

\begin{Center}
{{\sf\bfseries{\Huge{Review of Born--Infeld electrodynamics}}}

\bigskip

{Y.\,F.\,Alam\footnote{\url{yasin.alam@tamu.edu}} and A.\,Behne\footnote{\url{alexbehne@tamu.edu}}}}

\setcounter{footnote}{0}

\smallskip

{\itshape Department of Physics and Astronomy, Texas A\,\&\,M University \\ College Station, TX 77843, USA}
\end{Center}

\begin{abstract} \noindent \normalsize
        Born--Infeld electrodynamics is motivated by the infinite self-energy of the point charge in Maxwell electrodynamics. In \textsc{bi} electrodynamics, an upper bound $b$ is imposed on the electric field, thus limiting the self-energy of the point charge. 
        This is a review paper in which we motivate the \bi\ Lagrangian and from it derive the field equations. 
        We find the stress--energy tensor in \bi. 
        We calculate the potential due to the point charge in \bi. 
        We find order $b^{-2}$ wave solutions to \bi\ in $1+1$ dimensions. 
        We examine \bi\ plane waves normally incident on a mirror.
\end{abstract}



\setcounter{tocdepth}{2}
\tableofcontents

\section{Motivation} \label{sect:mot}

In \cite{BI34}, Born and Infeld describe their theory. In Maxwell electrodynamics, the Lagrangian is%
\begin{equation} \label{eq:maxL}
    \Lagr = - \frac{1}{4} F_{\mu\nu} F^{\mu\nu} = - \frac{1}{2}\left( E^2 - B^2 \right).
\end{equation}
The field due to a point charge is given by Coulomb's law which implies $E \to \infty$ at the position of the particle. This leads to an infinite Lagrangian and self-energy for the point particle. In order to eliminate this infinity the following replacement is proposed.
\begin{align} 
    \Lagr &= b^2\left(1-\sqrt{1-\frac{F_{\mu\nu} F^{\mu\nu}}{2 b^2}}\right)
    \nonumber
    \\
    &= b^2\left(1-\sqrt{1-\frac{E^2 - B^2}{b^2}}\right)
    \label{eq:BL}
\end{align}
where $b$ is some large constant. This is analogous to, in special relativity, replacing $mv^2/2$ with $mc^2(1-\sqrt{1-v^2/c^2})$. The immediate effect is to set an upper bound $b$ on $E$ just as $c$ is an upper bound on velocity. (\ref{eq:maxL}) is recovered from (\ref{eq:BL}) by Taylor expansion for $b^2 \gg F_{\mu\nu} F^{\mu\nu} $.
\eq{eq:BL} is only an ansatz, the Born--Infeld Lagrangian (in Minkowski space) is%
\begin{align}
    \Lagr &= b^2\left( 1- \sqrt{-\det{\left(\eta_{\mu\nu} + \frac{1}{b} F_{\mu\nu} \right)}} \right)
    \nonumber
    \\
    &= b^2 \left(1 - \sqrt{1-\frac{E^2-B^2}{b^2} - \frac{(\mathbf{E} \cdot \mathbf{B})^2}{b^4}} \right).
     \label{eq:BIL}
\end{align}
\eq{eq:BL} and \eq{eq:BIL} are equivalent for the static case. \eq{eq:BIL} is the simplest linear combination of 
$[-\det{(\eta_{\mu\nu}+F_{\mu\nu}/b)}]^{1/2}$,
$[-\det{(\eta_{\mu\nu})}]^{1/2}$, and 
$[-\det{(F_{\mu\nu}/b)}]^{1/2}$ 
(these expressions are all generally covariant) from which \eq{eq:maxL} is recovered upon Taylor expansion for $F_{\mu\nu} \ll b$. This is discussed further in section {\ref{sect:lag}}.

The infinite self-energy caused problems in the formulation of a quantum theory of electrodynamics. Born--Infeld is a 1930s attempt to unify quantum mechanics and field electrodynamics \cite{Goe14}. The theory does not achieve this goal (it is not the basis of quantum electrodynamics) but is interesting nonetheless. In the 1970s, it was discovered that Born-Infield is (except for a physically uninteresting case) the only nonlinear theory which does not predict vacuum birefringence and has correct shock wave characteristics \cite{Boi70,Boi72}. In the 1980s, string theorists found that the Born--Infeld action is also the action for an open superstring when the derivatives of the spacetime fields can be neglected; that is, for slowly varying fields \cite{Berg87,Met87}. 

In this paper, we review basic aspects of Born--Infeld electrodynamics, including the field equations \cite{BI34}, the stress--energy tensor\cite{BI34}, the electrostatic field due to a point charge \cite{BI34}, the wave equation in $1+1$ dimensions \cite{Man20}, standing wave solutions \cite{Man20}, free wave solutions \cite{Fer16}, and the particular standing waves which arise from plane waves normally incident on a mirror \cite{Fer16}.


\section{Lagrangian and equations of motion} \label{sect:lag} 

The principle underlying the Lagrangian postulated is that the action integral must be generally invariant. \cite{BI34} has the goal of finding the conditions that satisfy this postulate and under which \eq{eq:maxL} is recovered for small fields. The Lagrangian must have the form
\begin{equation}
    \Lagr = \sqrt{\det \left(a_{\mu \nu}\right)}
\end{equation}
where $a_{\mu\nu}$ is some covariant tensor. $a_{\mu \nu}$ can be split into an anti-symmetric tensor---we can use the electromagnetic tensor $F_{\mu \nu}$---and symmetric tensor---we can use the metrical tensor $g_{\mu \nu}$---such that $a_{\mu \nu}=g_{\mu \nu} + F_{\mu \nu}$. As the tensor is second order, $\Lagr$ can be any homogeneous function of the determinants of the tensors to the power of $1/2$. Therefore, following \cite{BI34} we consider the Lagrangian
\begin{equation}
    \Lagr = \sqrt{-\det \left(g_{\mu \nu}+F_{\mu \nu}\right)}+A\sqrt{-\det \left(g_{\mu \nu}\right)}+B\sqrt{\det \left(F_{\mu \nu}\right)}
\end{equation}
we have set $b=1$ to simplify the notation, $A$ and $B$ are some constants, and the negative signs have been added to keep $\L$ real. The last term is a total derivative that drops out of the action $S=\int \d^4 x\, {\cal L}$ since
\begin{align}
\sqrt{\det \left(F_{\mu \nu}\right)}=\frac{1}{4} \sqrt{-g} F_{\mu \nu} \tilde{F}^{\mu \nu} = \frac14 \partial_\mu \left(\epsilon^{\mu\nu\rho\sigma} A_\nu F_{\rho\sigma}\right)\ ,
\end{align}
where $g = \det g_{\mu\nu}$, $\tilde{F}^{\mu\nu} =\frac{1}{2\sqrt{-g}} \epsilon^{\mu\nu\rho\sigma} F_{\rho\sigma}$, and $\epsilon^{\mu\nu\rho\sigma}$ is the Levi-Civita symbol with $\epsilon^{0123} = 1$. Thus one can set $B=0$.
Furthermore, requiring that the Lagrangian gives the the classical Maxwell Lagrangian, for small values of $F_{\mu \nu}$, we must set $A=-1$.  This gives \cite{BI34} \footnote{Note that $\det \left(F_{\mu \nu}\right) = \left(\mathbf{B} \cdot \mathbf{E}\right)^2$ and $\frac{1}{2}F_{\mu \nu}F^{\mu \nu} = \left(\mathbf{B}^2 - \mathbf{E}^2\right)$.}
\begin{align}
{\cal L} &= \sqrt{-\det \left(g_{\mu \nu}+F_{\mu \nu}\right)}-\sqrt{-g}
\nn\\
 &= \sqrt{-g} \left(\sqrt{ 1+\frac12 F_{\mu \nu}F^{\mu \nu} - g^{-1} \det \left(F_{\mu\nu}\right)}-1\right)
 \label{eq:action}
\end{align}
where  $F_{\mu \nu}F^{\mu \nu} = g^{\mu\rho} g^{\nu\sigma} F_{\mu\nu} F_{\rho\sigma}$. This can be derived by writing 
\begin{equation}
\det \left(g_{\mu \nu}+F_{\mu \nu}\right)= \det \Big[g_{\mu\rho} \left(\delta^\rho_\nu + F^\rho{}_\nu \right)\Big] = g \det \left(\delta^\rho_\nu + F^\rho{}_\nu\right) =  g \det \left(\delta^\rho_\nu + g^{\rho \mu} F_\mu{}_\nu\right) \,
\end{equation}
and using the general formula valid for antisymmetric $4\times 4$ matrix $X$ given by
\begin{equation}
\det (1+X) = 1 - \frac12 \tr {X^2} + \det X\ . 
\end{equation}
This, in turn, can be derived from the general formula $\det Y = \exp {[ \tr {(\log Y)} ]}$ for a general matrix $Y$, easily seen by diagonalizing the matrix $Y$.

To find the field equations, we minimize the action by varying the potential as follows.
\begin{align}
    \delta S &= \int \d^4 x \, \delta_A \left[\sqrt{-g}\left(\sqrt{1+\frac{1}{2}F_{\mu \nu}F^{\mu \nu}- g^{-1} \det \left(F_{\mu \nu}\right)}-1\right)\right] \nonumber
    \\
    &=\int \d^4 x \left[\frac{\sqrt{-g}}{2\sqrt{\Pi}}\left(\frac{\dell \left(\frac{1}{2}F_{\mu \nu}F^{\mu \nu}\right)}{\dell F_{\mu \nu}} \delta F_{\mu \nu} -   \frac{\dell \left( g^{-1} \det \left(F_{\mu \nu}\right)\right)}{\dell F_{\mu \nu}}  \delta F_{\mu \nu}\right)\right] \nonumber \\
    &=\int \d^4 x \left[\frac{\sqrt{-g}}{\sqrt{\Pi}}\left(\frac{1}{2} F^{\mu \nu}  - 2 \left(\frac{1}{4}F_{\mu \nu} \tilde{F}^{\mu \nu}\right)  \frac{1}{4} \tilde{F}^{\mu \nu} \right)\dell_{\mu}\delta A_{\nu}\right] \label{eq:calcBI}
\end{align}
where 
$
    \Pi = 1+\frac{1}{2}F_{\mu \nu}F^{\mu \nu}- g^{-1} \det \left(F_{\mu \nu}\right).
$
Integrating by parts and dropping the surface term we find
\begin{equation}
    \delta S =\int \d^4x \, \dell_{\mu}\left[\frac{\sqrt{-g}}{\sqrt{\Pi}}\left(\frac{1}{2} F^{\mu \nu} - \frac{1}{2} \GG \tilde{F}^{\mu \nu}\right)\right] \delta A_{\nu}.
\end{equation}
Thus, the variational principle, $\delta S = 0$ gives the field equation
\begin{equation}
    \nabla_{\mu}\left[\frac{1}{\sqrt{\Pi}}\left(F^{\mu \nu} - \left(\frac{1}{4}F_{\alpha \beta} \tilde{F}^{\alpha\beta}\right) \tilde{F}^{\mu \nu}\right)\right] = 0 \label{eq:EOM}
\end{equation}
where the covariant derivative $\nabla_\mu$ is defined as
\begin{align}
    \nabla_\mu V^\nu &= \partial_\mu V^\nu + \Gamma_{\mu\rho}^\nu V^\rho\ ,
   \nonumber \\
    \nabla_\mu V_\nu &= \partial_\mu V_\nu -\Gamma_{\mu\nu}^\rho V_\rho\ ,
\end{align}
where $\Gamma_{\mu\nu}^\rho = \tfrac12 g^{\rho\sigma} \big[ \partial_\mu g_{\nu\sigma} + \partial_\nu g_{\mu\sigma}-\partial_\rho g_{\mu\nu} \big]$. Thus, we have
\begin{equation}
    \nabla_\mu V^{\nu\rho} = \partial_\mu V^{\nu\rho} -\Gamma_{\mu\sigma}^\nu V^{\sigma \rho} -\Gamma_{\mu\sigma}^\rho V^{\nu\sigma}.
\end{equation}
Reintroducing the parameter $b$, the Born--Infeld action takes the form
\begin{equation} \label{eq:bilagr}
    \Lagr = \sqrt{-g}\left(\sqrt{1 + \frac{1}{b^2}\left(\frac{1}{2} F_{\mu \nu}F^{\mu \nu}\right) - \frac{1}{g b^4}\det \left(F_{\mu \nu}\right)} -1 \right)b^2
\end{equation}
and the field equations are given by 
\begin{equation}
    \nabla_{\mu} \left[ \frac{1}{\sqrt{\Pi}} \left( F^{\mu \nu} - \frac{\GG}{b^2} \tilde{F}^{\mu \nu}\right)\right] = 0\ ,
\end{equation}
where 
\begin{equation}
\Pi = 1 + \frac{1}{b^2}\left(\frac{1}{2} F_{\mu \nu}F^{\mu \nu}\right) - \frac{1}{g b^4}\det \left(F_{\mu \nu}\right)\ .
\end{equation}
The electric displacement field given by $\mathbf{D} =b^2 \frac{\dell \mathbf{L}}{ \dell \mathbf{E}}$ and $H$-field $\mathbf{H} = b^2 \frac{\dell \mathbf{L}}{ \dell \mathbf{B}}$ are
\begin{align}
    \mathbf{D} = \frac{1}{\sqrt{\Pi}} \left(\mathbf{E} - \frac{\GG}{b^2} \mathbf{B}\right) \quad &\x{and} \quad \mathbf{H} = \frac{1}{\sqrt{\Pi}} \left(\mathbf{B} - \frac{\GG}{b^2} \mathbf{E}\right) \nonumber
    \\
    \x{or} \quad \mathbf{D} = \varepsilon\mathbf{E} - \nu \mathbf{B} \quad &\x{and} \quad
    \mathbf{H} = \frac{1}{\mu} \mathbf{B} - \nu \mathbf{E} \nonumber
    \\
    \with \varepsilon = \frac{1}{\sqrt{\Pi}}, \quad \mu &= \sqrt{\Pi},\quad \nu = \frac{\GG}{b^2 \sqrt{\Pi}}. \label{eq:constRelat}
\end{align}
These yield a nonlinear version of the Maxwell equations. From the equations above and the Bianchi identity \cite{BI34}
\begin{align}
    \mathbf{\nabla} \cdot \mathbf{D} &= 0, \nonumber\\
    \frac{\dell \mathbf{D}}{\dell t} &= \mathbf{\nabla} \times \mathbf{H}, \nonumber\\
    \mathbf{\nabla} \cdot \mathbf{B} &= 0, \nonumber\\
    -\frac{\dell \mathbf{B}}{\dell t} &= \mathbf{\nabla} \times \mathbf{E}. \label{eq:max}
\end{align}

\section{Stress--energy tensor}

 In order to find the stress--energy tensor, we must vary the action $S$ with respect to the metric $g_{\mu\nu}$. This will give us the following relation. 
\be \label{eq:stressEnergy}
    \delta S = -\frac12 \int \d^4 x  \sqrt{-g} T_{\mu\nu} \delta g^{\mu\nu}
\ee
where $g^{\mu\nu}$ is the inverse metric.
\begin{align}
    g^{\mu\nu} g_{\nu\rho} &= \delta^\mu_\rho 
    \nonumber
    \\
    \delta \sqrt{-g} &= \frac12 \sqrt{-g} \, g^{\mu\nu} \, \delta g_{\mu\nu} 
    \label{eq:varg}
\end{align}
Note that
$
g^{\mu\nu} \, \delta g_{\mu\nu} = -(\delta g^{\mu\nu} )g_{\mu\nu}.
$
Starting from \eq{eq:calcBI}, 
\begin{align}
    \delta S &= \int \d^4 x \, \delta_g\left[\sqrt{-g}\left(\sqrt{1+\frac{1}{2}F_{\mu \nu}F^{\mu \nu}- g^{-1} \det \left(F_{\mu \nu}\right)}-1\right)\right]
    \nonumber
    \\ 
    &= \int \d^4 x \left[ \left(\delta_g\sqrt{-g}\right)\mathbf{L} + \sqrt{-g} \, \delta_g \mathbf{L} \right]
    \label{eq:metricAction}
\end{align}
where $\mathbf{L} = \sqrt{1+\frac{1}{2}F_{\mu \nu}F^{\mu \nu}- g^{-1} \det \left(F_{\mu \nu}\right)}-1 $. Variation of $\sqrt{-g}$ is given by (\ref{eq:varg}). As for the variation of $\mathbf{L}$, first note
\begin{align}
    \frac{\delta}{\delta g^{\mu \nu}} \GGmu = \frac{\delta}{\delta g^{\mu \nu}} \left(\frac14 F_{\mu \nu}  \frac{1}{2\sqrt{-g}} \epsilon^{\mu\nu\rho\sigma} F_{\rho\sigma} \right) &= \frac12 \GG g_{\mu \nu} 
    \\
    \frac{\delta}{\delta g^{\mu \nu}} \left(\frac{1}{2}F_{\mu \nu}F^{\mu \nu}\right) =  \frac{\delta}{\delta g^{\mu \nu}} \left(\frac{1}{2}F_{\mu \rho}F_{\nu \sigma} g^{\mu \nu} g^{\rho \sigma} \right) &= 2 \left(\frac12 F_{\mu \rho} g^{\rho \sigma} F_{\nu \sigma}\right)
\end{align}
as, by definition, $F^{\mu \nu}$ is independent of the metric. From here, it is easy to see
\begin{align}
    \delta_g \mathbf{L} &= \frac{1}{2\sqrt{\Pi}}\left(F_{\mu \rho} g^{\rho \sigma} F_{\nu \sigma} - \GG^2 g_{\mu \nu}\right)\delta g^{\mu\nu}.
\end{align}
So (\ref{eq:metricAction}) can be rewritten 
\begin{align}
    \delta S &= \int d^4 x \left[ \left(-\frac12 \sqrt{-g} g_{\mu\nu} \delta g^{\mu\nu}\right)\mathbf{L} \right.
    \nn\\
    &\qquad+ \left. \sqrt{-g} \frac{\delta g^{\mu\nu}}{2\sqrt{\Pi}}\left(F_{\mu \rho} g^{\rho \sigma} F_{\nu \sigma} - \GG^2 g_{\mu \nu}\right) \right].
\end{align}
From this, (\ref{eq:stressEnergy}), and the necessity that the functional derivative must coincide with 
$
    -\sqrt{-g}/2 T_{\mu \nu}
$
we obtain \cite{BI34}
\begin{align}
    T_{\mu \nu} &= \frac{-2}{\sqrt{-g}}\frac{\delta S}{\delta g^{\mu\nu}}
    \nn\\
    &= g_{\mu\nu}\mathbf{L} - \frac{1}{\sqrt{\Pi}}\left(F_{\mu \rho} F_\nu{} ^ {\rho} - \GG^2 g_{\mu \nu}\right).
\end{align}

\section{Point charge}

The electric field due to a point charge in Born--Infeld electrodynamics was derived in \cite{BI34}. 
Following \cite{BI34} we start by considering a stationary point charge $q$ at the origin. The problem is static ($\mathbf{B} = \mathbf{H} = 0$ and nothing depends on $t$) so \eq{eq:max} becomes
\begin{align}
     \quad \nabla \times \mathbf{E} &= 0, \nonumber
     \\
     \nabla \cdot \mathbf{D} &= 0. \label{eq:divD}
\end{align}
And \eq{eq:constRelat} is
\begin{align}
    \mathbf{D} = \frac{\mathbf{E}}{\sqrt{1-\frac{E^2}{b^2}}}.
    \label{eq:pcDEconst}
\end{align}
Also, since $\mathbf{E}$ is curl free, $\mathbf{E}=-\nabla \Phi$ where $\Phi$ is some scalar potential.

The problem is spherically symmetric and static so the fields must depend only on $r$ and be only in the $\rhat$ direction. So \eq{eq:divD} is nothing but%
\begin{align}
    \frac{\d}{\d r} r^2 D_r &= 0 \nonumber
    \\
    D_r &= \frac{C}{r^2}
\end{align}
where $C$ is some constant which we can determine by applying Gauss's law to a sphere centered at the origin.
\begin{align}
    \oint \mathbf{D} \cdot \d\mathbf{a} &= q \nonumber
    \\
    4 \pi C &= q.
\end{align}
So $D_r = (q) / (4 \pi r^2)$. Now from this result, \eq{eq:pcDEconst}, and $E_r = -\partial \Phi / \partial r = -\d \Phi / \d r$ we have the following equation.
\begin{align}
    \frac{q}{4 \pi r^2} = -\frac{\d \Phi / \d r}{\sqrt{1 - \frac{1}{b^2} \left(\d \Phi / \d r\right)^2}}.
\end{align}
This is separable.
\begin{align}
    (\d \Phi)^2 &= \left(\frac{q}{4 \pi r_0^2}\right)^2 \frac{(\d r)^2}{1 + (r/r_0)^4} \nonumber \\
    \d \Phi &= - \frac{q}{4 \pi r_0^2} \frac{\d r}{\sqrt{1+(r/r_0)^4}} \label{eq:dPhiPoint}
\end{align}
where the substitution $r_0^2 := |q| / (4 \pi b)$ has been made and the overall sign is determined by the physical condition that the signs of $\d \Phi / \d r$ and $q$ differ. We take the reference point to be $r \to \infty$ and integrate \eq{eq:dPhiPoint} from the reference point to $r$ which, along with the substitution $u := r/r_0$, yields
\begin{align}
    \Phi = \frac{q}{4 \pi r_0} f\left(\frac{r}{r_0}\right) \quad \text{where} \quad f(x) := \int_x^\infty \frac{\d u}{\sqrt{1 + u^4}}.
\end{align}
This is the potential of the point charge in Born--Infeld. For $x \gg 1$ (so for large $b$ or large $r$), Coulomb's law is recovered.
 
Making the substitution $u = \tan{\beta / 2}$ leads to \cite{BI34}
\begin{align}
     f(x) 
     &= \frac{1}{2} \int_{\beta'}^\pi \frac{\d \beta}{\sqrt{1 - \frac{1}{2}\sin^2{\beta}}} 
     = \frac{1}{2} \int_{0}^\pi \frac{\d \beta}{\sqrt{1 - \frac{1}{2}\sin^2{\beta}}} 
     - \frac{1}{2} \int^{\beta'}_0 \frac{\d \beta}{\sqrt{1 - \frac{1}{2}\sin^2{\beta}}} \nonumber
     \\
     &= f(0) - \frac{1}{2} F\left(\beta', \frac{1}{\sqrt{2}}\right) 
\end{align}
where $\beta' = 2 \arctan{x}$ and $F(\beta', 1/\sqrt{2})$ is the elliptic integral of the first kind for coefficient $1/2$ and upper bound $\beta'$.

To find $f(0)$ we evaluate the elliptic integral 
\begin{align}
    f(0) = \frac{1}{2} \int_{0}^\pi \frac{\d \beta}{\sqrt{1 - \frac{1}{2}\sin^2{\beta}}} = \frac{1}{2} F \left(  \pi, \frac{1}{\sqrt{2}} \right) = 1.85.
\end{align}
This is the maximum value of $f$ and so the maximum value of $\Phi$ is $(1.85\times q) / (4 \pi r_0)$ at $r=0$.

\begin{SCfigure}
	\includegraphics[width = 0.7\linewidth]{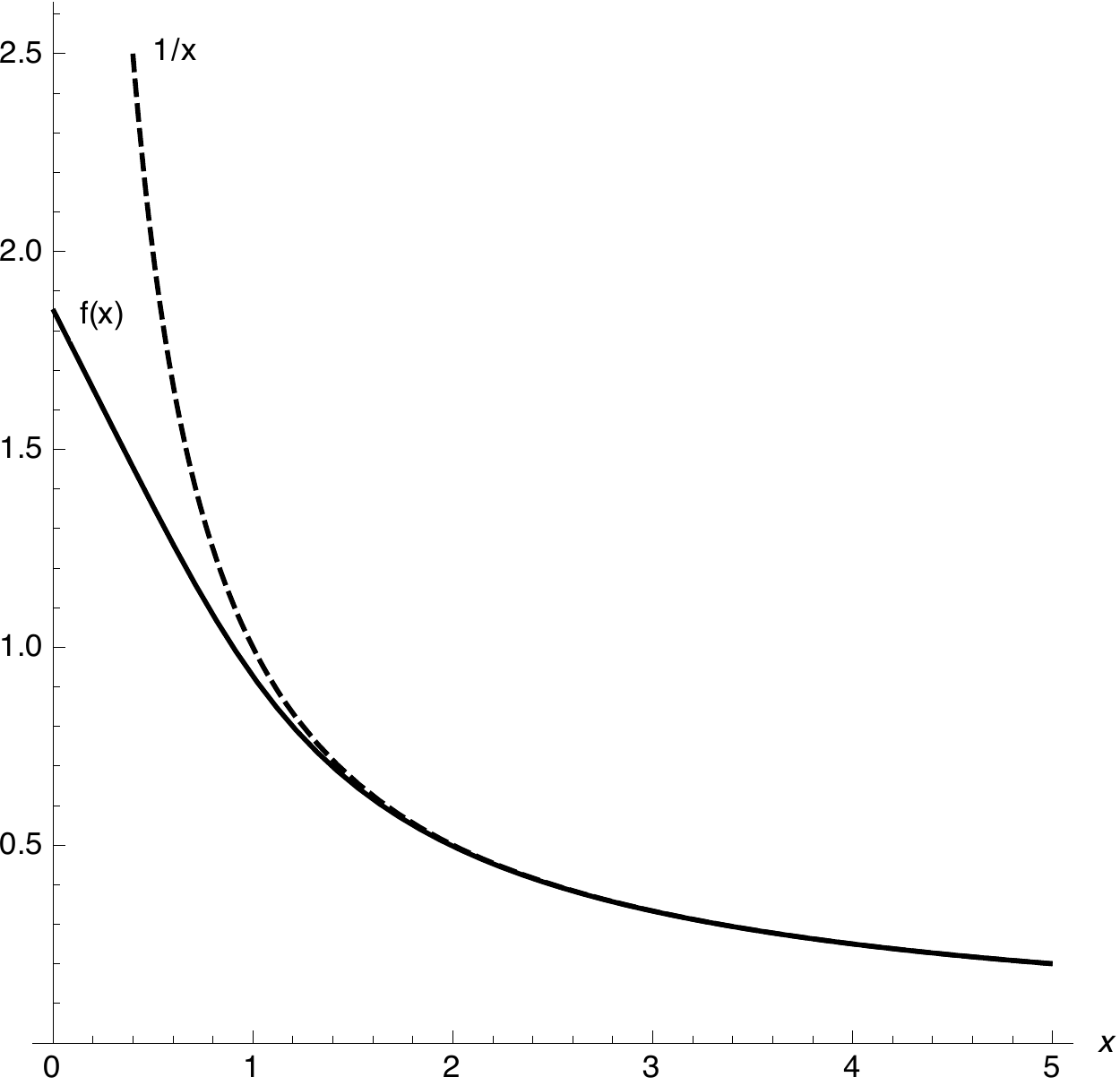}
	\caption{The potentials of the point charge in Born--Infeld (solid) and Maxwell electrodynamics (dashed).}
	\label{fig:point}
\end{SCfigure}

The Born-Infield and Maxwell (Coulomb) potentials for a point charge are compared in figure \ref{fig:point}.

From \eq{eq:dPhiPoint}
\begin{align}
    \mathbf{E} = - \rhat \frac{\d \Phi}{\d r} = \frac{q}{4 \pi r_0^2} \frac{\rhat}{\sqrt{1+(r/r_0)^4}}.
\end{align}

So in Born--Infeld electrodynamics, the potential and the electric field of the point charge are finite everywhere and have a discontinuity at the position of the point charge.

\section[Wave equation in 1 + 1 dimensions and its solutions]{Wave equation in 1\;+\;1 dimensions and its solutions}

In this section, following \cite{Man20}, we wee will review the derivation of a $1+1$ dimensional wave equation for the electromagnetic field in Born--Infeld electrodynamics. We will use the Poincar{\'e}-Lindstedt method to find a solution for a standing wave in $1+1$ dimensions in between two infinite plates. We will also look at a free wave solution in $1+1$ dimensions to $\mathcal{O} (b^{-4})$.

\subsection{Wave equation}

Let us consider an electromagnetic wave propagating in the $x$ direction and polarized in the $y$ direction. To this end, and choosing the gauge $A_z=0$, we consider following ansatz
\begin{align}
    A_y\left(\mathbf{r},t\right) = u\left(x,t\right), \quad
    \Phi = A_x = A_z = 0
\end{align}
which gives 
\begin{align}
    F_{02}=E_y=-\dell_t A_y &= -u_t
    \label{eq:Ey}
    \\
    \and F_{12}=-B_z=-\dell_x A_y &= -u_x
    \label{eq:Bz}
\end{align}
where we adopt the convention $u_x := \dell u / \dell x$. A defense of why this is a good choice of gauge can be found in \cite{Man20}. We can put (\ref{eq:Ey}) and (\ref{eq:Bz}) into (\ref{eq:EOM}) with the condition of flat spacetime. It can be seen that, in flat spacetime, $\frac{1}{4}F_{\alpha \beta} \tilde{F}^{\alpha \beta} = 0$ which simplifies (\ref{eq:EOM}) to 
\begin{align}
    \dell_u\left(\frac{1}{\sqrt{\Pi}} F^{\mu \nu}\right) &= 0
    \\
    \dell_t \left(\frac{1}{\sqrt{\Pi}} F^{0 2}\right) + \dell_x \left(\frac{1}{\sqrt{\Pi}} F^{1 2}\right) &= 0.
\end{align}
It is useful to note
\begin{align}
    \Pi =  1 + b^{-2}\left(\frac12 F_{\mu \nu} F^{\mu \nu}\right) = 1 + b^{-2}\left(u_x^2-u_t^2\right).
\end{align}
The equations of motion simplify to (by a hearty application of the chain rule)
\begin{align}
    \left(1-\frac{1}{b^2}u_t^2\right)u_{xx} - \left(1+\frac{1}{b^2}u_x^2\right)u_{tt}+\frac{2}{b^2}u_x u_t u_{xt}=0.
    \label{eq:wave}
\end{align}
Apart from $1+1$, it is easy to convert (\ref{eq:wave}) to 2 dimensional form for $\mathbf{B} = 0$ with the substitution $t \rightarrow iy$ to obtain the equation \cite{Man20}
\begin{align}
    \left(1-\frac{1}{b^2}u_y^2\right)u_{xx} - \left(1-\frac{1}{b^2}u_x^2\right)u_{yy}+\frac{2}{b^2}u_x u_y u_{yx}=0.
    \label{eq:wave2d}
\end{align}
(\ref{eq:wave}) is the equation of motion for an electromagnetic wave. In the next section, we will review a solution that describes standing wave between two parallel conducting plates. We will also review an approximate wave solution.

\subsection{Standing wave solution}

In this section, following \cite{Man20}, we describe standing wave solution between two infinite parallel conducting plates. Let us take these plates to be in the $y$--$z$ plane at $x=0$ and $x=L$ at which the electric field must vanish. The other condition (which ensures the solutions are oscillatory) is 
\begin{align}
    1+b^{-2}\left(u_x^2-u_t^2\right) > 0.
\end{align}
We will use the iterative Poincar{\'e}-Lindstedt method to solve \eq{eq:wave} up to higher and higher orders of $b^{-1}$. \eq{eq:wave} can be rewritten
\begin{align}
    u_{xx} - u_{tt} - b^{-2} ( u_t^2 u_{xx} + u_x^2 u_{tt} - 2 u_x u_t u_{xt} ) &= 0. \label{eq:wave*}
\end{align}
We neglect $\ordrb{2}$ terms and get $u_{xx} - u_{tt} = 0$ which has the satisfactory solution 
\be
    u^{(0)} = A \sin{kx} \cos{\omega t} = \frac{A}{2} [ \sin{(kx + \omega t)} + \sin{(kx - \omega t)} ].
\ee
We define 
\be
s_{nm} := \sin(n k x + m \omega t) + \sin{(n k x - m \omega t)}
\ee
so that
$
    \usol{0} = \frac{A}{2} s_{11}.
$
We evaluate the \textsc{lhs} of \eq{eq:wave*} with $u = \usol{0}$ and get
\be
    \frac{A}{2} ( \omega^2 - k^2 ) s_{11} - \frac{A^3 \omega^2 k^2}{8 b^2} ( s_{13} - s_{31} - 2 s_{11} ).
\ee
We want this expression to be 0. We'll look at each coefficient on each $s_{mn}$. First, for the coefficient on $s_{11}$ to be 0, we can impose the following condition on the frequency.
\begin{align}
    \omega^2 - k^2 + \frac{\omega^2 \epsilon^2}{2} = 0 \where \epsilon := \frac{A k}{b}. \label{eq:seedCoef}
\end{align}
We solve this for the frequency and Taylor expand in $\epsilon^2$ to get
\begin{align}
    \frac{\omega^2}{k^2} = 1 - \frac{\epsilon^2}{2} + \ordrep{4}. \label{eq:omega1}
\end{align}
Now what about the coefficients on $s_{13}$ and $s_{31}$? Since
\be
    (\dell_{x}^2 - \dell_{t}^2) s_{nm} = (m^2 \omega^2 - n^2 k^2) s_{nm},
\ee
we can eliminate those coefficients by adding
\begin{align}
    \usol{1} := \frac{A \epsilon^2}{8} \left( \frac{s_{13}}{9 \omega^2 - k^2} - \frac{s_{31}}{\omega^2 - 9k^2} \right) \label{eq:u1}
\end{align}
to $\usol{0}$. Since \eq{eq:omega1}, we can Taylor expand \eq{eq:u1} in $\epsilon$ to get
\be
    \usol{1} = \frac{A \epsilon^2}{64} (s_{13} + s_{31} ) + \ordrep{4}.
\ee

To go to the next order we again evaluate the \textsc{lhs} of \eq{eq:wave*} but with 
$
u = \usol{0} + \usol{1}.
$
and to $\ordrep{4}$. We write the coefficient on $s_{11}$, say it must be 0, solve for $\omega^2 / k^2$, and Taylor expand to $\ordrep{4}$. We look at all the $s_{n \neq m}$ terms and find $\usol{3}$.

This can be repeated for higher orders of $\epsilon$. It is not known whether the series converge.

\subsection{Free wave solution}

Next, following \cite{Fer16}, we describe free wave solution of \eq{eq:wave}.
It is convenient to work in the complex plane such that
\begin{align}
    \d x^2-\d t^2 = |\d z|^2 = \d z \id\ol{z}
\end{align}
and the metric tensor will have the components
\begin{align}
    g_{zz} = g_{\ol{zz}} = 0, \quad g_{z\ol{z}} = g_{\ol{z}z} = \tfrac12.
\end{align}

It is then possible to replace $u_x^2+u_y^2 = g^{\mu \nu}u_{\mu}u_{\nu} = 4 u_z u_{\ol{z}}$ which leads to a reformulation of (\ref{eq:wave}) as
\begin{align}
    u_{z\ol{z}} + \frac{1}{b^2} u_{\ol{z}}^2u_{zz} - \frac{2}{b^2} u_z u_{\ol{z}} u_{z\ol{z}} + \frac{1}{b^2} u_{z}^2 u_{\ol{zz}} = 0.
    \label{eq:wavez}
\end{align}

We can use the equipotential spacetime lines $u(x,t) = constant$ as a non-Cartesian basis and change coordinates $(z,\ol{z}) \rightarrow (w,\ol{w})$ where $w = u + iv$ and $(u,v) \in \mathbb{R}^2$. The Jacobian matrix is
\begin{equation}
    \begin{pmatrix}
        u_z & u_{\ol{z}} 
        \\
        v_z & v_{\ol{z}}
    \end{pmatrix}
    =
    \begin{pmatrix}
        \dpartialfrac{z}{u} & \dpartialfrac{z}{v}
        \\
        \dpartialfrac{\ol{z}}{u} & \dpartialfrac{\ol{z}}{v}
    \end{pmatrix} ^ {-1}
    =
    \begin{pmatrix}
        (z_w + z_{\ol{w}}) & i(z_w + z_{\ol{w}})
        \\
        (\ol{z}_w + \ol{z}_{\ol{w}}) & i(\ol{z}_w + \ol{z}_{\ol{w}})
    \end{pmatrix}^{-1}.
\end{equation}

At this point, it is worth revisiting the action. From (\ref{eq:action}) we have
\begin{equation}
    S[u] = \frac{b^2}{8 \pi i}\int dz \wedge d\ol{z} \sqrt{1-4b^{-2}u_{z}u_{\ol{z}}}.
    \label{eq:actionU}
\end{equation}
Note that (\ref{eq:wavez}) can be obtained from (\ref{eq:actionU}). It is easier to solve (\ref{eq:wavez}) for $z(w,\ol{w})$ and not $u(z,\ol{z})$ as it is possible to choose a convenient coordinate $v$. Rewriting the new action in terms of $z(w,\ol{w})$, we have
\begin{equation}
    dz \wedge d\ol{z} = (z_w \ol{z}_{\ol{w}} - z_{\ol{w}}\ol{z}_w) dw \wedge d\ol{w}
\end{equation}
and we can solve for $u_z$ to get
\begin{equation}
        u_z = \frac{1}{2} \frac{\ol{z}_w - \ol{z}_{\ol{w}}}{z_{\ol{w}} \ol{z}_w - z_w \ol{z}_{\ol{w}}}.
\end{equation}
So (\ref{eq:actionU}) can be rewritten
\begin{align}
    &dz \wedge d\ol{z} \sqrt{1-4b^{-2}u_{z}u_{\ol{z}}}
    \nonumber
    \\
    &\qquad = dw \wedge d\ol{w} \sqrt{(z_{\ol{w}} \ol{z}_w - z_w \ol{z}_{\ol{w}})^2 - b^{-2}(\ol{z}_w - \ol{z}_{\ol{w}})(z_{\ol{w}}-z_w)}.
    \label{eq:LDw}
\end{align}
To simplify (\ref{eq:LDw}), we choose $w = u(z,\ol{z}) + i v(z,\ol{z})$ such that the vector $\frac{\partial}{\partial w}$ is in the form
\begin{equation}
    2b\frac{\partial}{\partial w} = \xi \frac{\partial}{\partial z} + \frac{\partial}{\partial \xi}\frac{\partial}{ \partial \ol{z}}
\end{equation}
for some function $\xi$. Therefore
\begin{align}
    \dpartialfrac{z}{w} = \frac{\xi}{2b}, \quad \dpartialfrac{z}{\ol{w}} = \frac{1}{2b\ol{xi}}.
    \label{eq:holo}
\end{align}
This implies $(u,v)$ are orthogonal. (\ref{eq:LDw}) becomes
\begin{equation}
    \frac{dw \wedge d\ol{w}}{4b^2}\left(\frac{1}{|\xi|^2}-2+ |\xi|^2 \right)=dw \wedge d\ol{w}\left(\left|\dpartialfrac{z}{\ol{w}}\right|^2 - 2 + \left|\dpartialfrac{z}{w}\right|^2 \right) 
    \label{eq:LDxi}
\end{equation}
which leads to the equations of motion
\begin{equation}
    \frac{\partial^2 z}{\partial w \, \partial \ol{w}} = 0.
    \label{eq:EOMW}
\end{equation}
Note that although $b$ is absent in (\ref{eq:EOMW}), the derivatives of $z(w,\ol{w})$ are connected to (\ref{eq:holo}).

This gives the solution (with a factor $4b^2$ for convenience) of
\begin{equation}
    z = f(w) + \frac{g(\ol{w})}{4 b^2}
    \label{eq:implicitsol}
\end{equation}
and where $f$ and $g$ are linked by 
\begin{equation}
    \ol{f'(w)} \, g'(\ol{w}) = 1.
    \label{eq:linkingcondition}
\end{equation}

Our task then is to invert \eq{eq:implicitsol} to find $w(z,\ol{z})$ of which the real part is the potential $u(z,\ol{z})$. We do this perturbatively.

To begin, we differentiate \eq{eq:implicitsol} with respect to $z$ and then $\ol{z}$ to get
\begin{align}
    1 &= f'(w) \partialfrac{w}{z} + \frac{g'(\ol{w})}{4 b^2} \partialfrac{\ol{w}}{z},
    \label{eq:zpartl}
    \\ 
    0 &= f'(w) \partialfrac{w}{\ol{z}} + \frac{g'(\ol{w})}{4 b^2} \partialfrac{\ol{w}}{\ol{z}}
    \label{eq:zbarpartl}
\end{align}
respectively. Solve \eq{eq:zbarpartl} for $\fpartialfrac{w}{\ol{z}}$, conjugate to get $\fpartialfrac{\ol{w}}{z}$, and substitute into \eq{eq:zpartl}. This yields
\begin{equation}
    1 = f'(w) \partialfrac{w}{z} - \frac{1}{16b^4} \frac{|g'(\ol{w})|^2}{\ol{f'{w}}} \partialfrac{w}{z} = f'(w) \partialfrac{w}{z}.
    \label{eq:1fprime}
\end{equation}
at the first order in $b^{-2}$. Furthermore, with this $\fpartialfrac{\ol{w}}{z}$, the $z$ derivative of \eq{eq:implicitsol} becomes
\begin{align}
    1 &= f'(w) \partialfrac{w}{z} + \ordr{b}{-2 \times 2}
    \nonumber
    \\
    f'(w) &= \partialfrac{z}{w}.
    \label{eq:fprimezw1}
\end{align}
Substituting this into \eq{eq:1fprime} yields
\begin{equation}
    \partialfrac{z}{w} \partialfrac{w}{z} = 1.
    \label{eq:highlynontrivial}
\end{equation}
This is trivial only in Maxwell electrodynamics where $w$ is analytic in $z$. In \textsc{bi} electrodynamics, $w$ is a function of $z$ and $\ol{z}$. But \eq{eq:highlynontrivial} says what is trivial in Maxwell is true in \textsc{bi}; at least, at the first order. From \eq{eq:linkingcondition} and \eq{eq:1fprime} we find
\begin{equation}
    g'(\ol{w}) = \frac{1}{\ol{f'(w)}} = \partialfrac{\ol{w}}{\ol{z}}.
\end{equation}
Substitute this and \eq{eq:fprimezw1} into \eq{eq:zbarpartl} to get
\begin{align}
    \partialfrac{w}{\zbar} + \frac{1}{4b^2} \partialfrac{w}{z} \left( \partialfrac{\ol{w}}{\zbar} \right)^2 = 0.
    \label{eq:wdiffeq}
\end{align}

Now we have only to solve \eq{eq:wdiffeq} for $w$ and take the real part. Drawing inspiration from the Maxwell solution $w_\x{M}(z) = D \cos{kz}$, let us take as ansatz
\begin{equation}
    w(z,\ol{z}) = D \cos{[k z + \Psi(\ol{z})]} \quad k \in \mathbb{R}.
\end{equation}
Now substitute this into \eq{eq:wdiffeq} but in the term which is diminished by $b^{-2}$, use $w_\x{M}$ instead.
\begin{equation}
    \Psi'(\ol{z}) + \frac{k^3 \ol{D}^2}{4 b^2} \sin^2 k\ol{z} = 0
\end{equation}
from which we find
\begin{equation}
    \Psi(\ol{z}) = -\frac{k^3 \ol{D}^2}{8 b^2} \ol{z} + \frac{k^2 \ol{D}^2}{16 b^2} \sin{2k\ol{z}}.
\end{equation}
So then the ansatz is now
\begin{align}
    w(z,\ol{z}) &= D \cos{\left( kz -\frac{k^3 \ol{D}^2}{8 b^2} \ol{z} + \frac{k^2 \ol{D}^2}{16 b^2} \sin{2k\ol{z}} \right)}
    \nonumber
    \\
    &= D \cos{\left( kz - \frac{k^3 \ol{D}^2}{8 b^2} \ol{z} \right)} - \frac{k^2 D \ol{D}^2}{16 b^2} \sin{( 2 k \ol{z} )} \sin{\left( kz - \frac{k^3 \ol{D}^2}{8 b^2} \ol{z} \right)}.
    \label{eq:problematicw}
\end{align}
Note that in the limit $k^2 D^2 \ll b^2$, $w \to w_\x{M}$. But this solution is problematic; when the \textsc{lhs} of \eq{eq:wdiffeq} is evaluated with this solution, we get a second order term linear in $z$. So \eq{eq:problematicw} is only a small $|z|$ approximation. We can fix this by replacing
\be
    \ol{z} \to \ol{z} + \frac{\alpha}{b^2} z.
\ee
This is permissible since $\fpartialfrac{w}{z}$ only appears in \eq{eq:wdiffeq} in the term diminished by $b^{-2}$. With $\alpha = - D^2 k^2 / 8$ the solution becomes
\begin{align}
    w(z,\ol{z}) &= D \cos{\left[ \left( 1 + \frac{k^4 |D|^4 }{64 b^2} \right) kz - \frac{k^3 \ol{D}^2}{8 b^2} \ol{z} \right]}
    \nonumber
    \\
    &\qquad- \frac{k^2 D \ol{D}^2}{16 b^2} \sin{\left[ 2 k \left( \ol{z} - \frac{k^2 D^2}{8 b^2} z \right) \right]} \sin{\left[ kz - \frac{k^3 \ol{D}^2}{8 b^2} \ol{z} \right]}.
    \label{eq:satisfactoryw}
\end{align}
This solution is satisfactory; that is, the $b^{-4}$ terms in the \textsc{lhs} of \eq{eq:wdiffeq} for \eq{eq:satisfactoryw} are all bounded.

Now we have only to take the real part. Let $D = D_1 + i D_2$. Consider that, with $z = x + iy$
\be
    \x{Re}(w_\x{M}) = u_\x{M} = D_1 \cos{kx} \cos{iky} + i D_2 \sin{kx} \sin{iky}.
\ee
If we make the replacement $y \to -it$ and set $D_1 = a_1 + a_2$ and $D_2 = i ( a_2 - a_1 )$ we get
\be
    u_\x{M} = a_1 \cos{k(x+t)} + a_2 \cos{k(x-t)};
\ee
two waves of different amplitudes propagating in opposite directions. Let us do the same for \eq{eq:satisfactoryw}. The result is
\begin{align}
    u &= \left\{ a_1 \cos{\left[ \left( 1 - \frac{a_2^2 k^2}{2 b^2} \right) k x -  \left( 1 + \frac{a_2^2 k^2}{2 b^2} \right) k t \right]} \right.
    \nonumber
    \\
    &\qquad+ \left. a_2 \cos{\left[ \left( 1 - \frac{a_1^2 k^2}{2 b^2} \right) k x + \left( 1 + \frac{a_1^2 k^2}{2 b^2} \right) k t \right]} \right\}
    \nonumber
    \\
    &\qquad\times \left\{ 1 - \frac{a_1 a_2 k^2}{2 b^2} \sin{\left[ \left( 1 - \frac{a_2^2 k^2}{2 b^2} \right) k x -  \left( 1 + \frac{a_2^2 k^2}{2 b^2} \right) k t \right]} \right.
    \nonumber
    \\
    &\qquad\times \left. \sin{\left[ \left( 1 - \frac{a_1^2 k^2}{2 b^2} \right) k x + \left( 1 + \frac{a_1^2 k^2}{2 b^2} \right) k t \right]} \right\}. 
    \label{eq:wavesol}
\end{align}
When the \textsc{lhs} of \eq{eq:wave} is evaluated with \eq{eq:wavesol}, all the $b^{-4}$ order terms are bounded. This is the wave solution.

\subsection{Mirror}

Suppose a mirror is placed in the $y$--$z$ plane. Then $E_y = \partial u / \partial t$ must be zero at the mirror which is at $x=0$. This is the case for $a_1 - a_2 =: a$. \eq{eq:wavesol} then becomes
\begin{align}
    u &= 2 a \sin{\vartheta} \sin{\varphi} \left[ 1 + \frac{a^2 k^2}{4 b^2} ( \cos{2\vartheta} - \cos{2\varphi} ) \right]
    \nn\\
    &= 2 a \left\{ \sin{\vartheta} \sin{\varphi} + \frac{a^2 k^2}{8 b^2} [ \sin{\varphi} (\sin{3\vartheta} - \sin{\vartheta}) - \sin{\vartheta} ( \sin{3 \varphi} - \sin{\varphi} ) ] \right\} \label{eq:mirrorpot}
\end{align}
where
\begin{align}
    \vartheta := \left( 1 + \frac{a^2 k^2}{2 b^2} \right) k t
    \and
    \varphi := \left( 1 - \frac{a^2 k^2}{2 b^2} \right) k x.
\end{align}
Thus, the solution is a stationary wave solution given by $E_y=0$ and $u$ given in \eq{eq:mirrorpot}, with  the stationary wave nodes at $x_n$, derived from $\sin{\varphi|_{x_n}}=0$, given by

\begin{align} \left( 1 - \frac{a^2 k^2}{2 b^2} \right) k x_n &= n \pi.
\end{align}

So, at the order $b^{-2}$, the nodes depend on the amplitude and the third harmonic is present. Both the amplitude dependence and the third harmonic are diminished by a factor of $b^{-2}$.

\printbibliography

\end{document}